\documentclass[twocolumn, longbib]{aastex7}
\usepackage{graphicx}
\usepackage[inkscapelatex=false]{svg}
\usepackage[utf8]{inputenc}
\usepackage[T1]{fontenc}
\usepackage{mathptmx}
\usepackage{etoolbox}
\usepackage{amssymb,amsmath}
\usepackage{multirow}
\usepackage{url}
\def\UrlBreaks{\do\/\do-}
\usepackage{xcolor}
\usepackage{soul}

\begin{document}

\title{Observations of [\ion{O}{1}] emission in Comets C/2014 Q2 (Lovejoy) and C/2007 N3 (Lulin): Possible Influence of Solar Activity on Oxygen Line Ratios}

\author[0009-0005-4186-4424]{Ella J. Mayfield}
\affiliation{Department of Physics and Astronomy, Appalachian State University, 525 Rivers St., Boone, NC 28607, USA}
\email[show]{mayfieldej@appstate.edu}

\author[0000-0002-0622-2400]{Adam J. McKay}
\affiliation{Department of Physics and Astronomy, Appalachian State University, 525 Rivers St., Boone, NC 28607, USA}
\email[show]{mckayaj@appstate.edu}

\author[0000-0002-6702-7676]{Michael S. P. Kelley}
\affiliation{Department of Astronomy, University of Maryland, 4296 Stadium Dr., College Park, MD 20742, USA}
\email{msk@astro.umd.edu}

\author[0000-0003-4828-7787]{Anita L. Cochran}
\affiliation{Department of Astronomy, University of Texas at Austin, 2515 Speedway, Austin, Texas, 78712, USA}
\altaffiliation{McDonald Observatory}
\email{anita@astro.as.utexas.edu}

\begin{abstract}

Observing [\ion{O}{1}] emission to calculate an "oxygen line ratio" has been proposed as a potential proxy for direct CO$_2$ measurement in comets. However, the photochemistry governing [\ion{O}{1}] release into the coma is not well understood, and using theoretical release rates often yields different results than using empirical release rates determined in conjunction with direct space-based measurements of CO$_2$. We hypothesize that the accuracy of the release rates could depend on the level of solar activity at the time the comet is observed, which will be influenced by the solar cycle. We present observations and analysis of [\ion{O}{1}] emission in one comet observed near solar maximum, C/2014 Q2 (Lovejoy), and one near solar minimum, C/2007 N3 (Lulin). Our [\ion{O}{1}] measurements were obtained using two high spectral resolution optical spectrographs: the Tull Coud\'e spectrometer at McDonald Observatory and the ARCES spectrometer at Apache Point Observatory. We use empirical and theoretical models for [\ion{O}{1}] emission from the literature to derive multiple sets of inferred CO$_2$ abundances for these comets and compare to contemporaneous space-based measurements of CO$_2$. We find that the empirical model, which was developed based on comet observations obtained near solar maximum, reproduces the directly measured CO$_2$ abundances better for Lovejoy. Neither model accurately reproduces the direct measurement for Lulin. We discuss the implications of our findings for the accuracy and dependencies of the oxygen line ratio method for inferring CO$_2$ abundances in cometary comae.

\end{abstract}


\section{Introduction}
\label{sect:intro}

Comets formed during the earliest stages of the solar system, and have undergone very little heating or processing since they were formed. This means their composition provides a unique window into the physical and chemical conditions present during the early era of solar system formation. Specifically, studying the volatile composition of comets is crucial to understanding both current cometary activity and evolutionary processes. The primary volatiles present in most comets are H$_2$O, CO$_2$, and CO, and these species are widely considered the drivers of cometary activity. Therefore, studying the sublimation behavior of these species is paramount to understanding cometary activity.

Both H$_2$O and CO are directly observable from the ground using near-infrared measurements, and CO is also observable in the sub-millimeter. However, the only direct method for measuring CO$_2$ is through the $v_3$ band at $4.26~\mu m$, which can only be observed using space-based facilities due to severe telluric absorption at this wavelength. Observing time on space-based facilities is often limited, so a ground-based proxy for direct CO$_2$ measurement is of extreme importance. Since atomic oxygen is a photodissociation product of H$_2$O, CO$_2$, and CO, observations of the forbidden oxygen lines at 5577, 6300, and 6364\AA~can serve as a proxy for direct measurement of these species.

\begin{figure}
    \centering
    \includegraphics[width=\columnwidth]{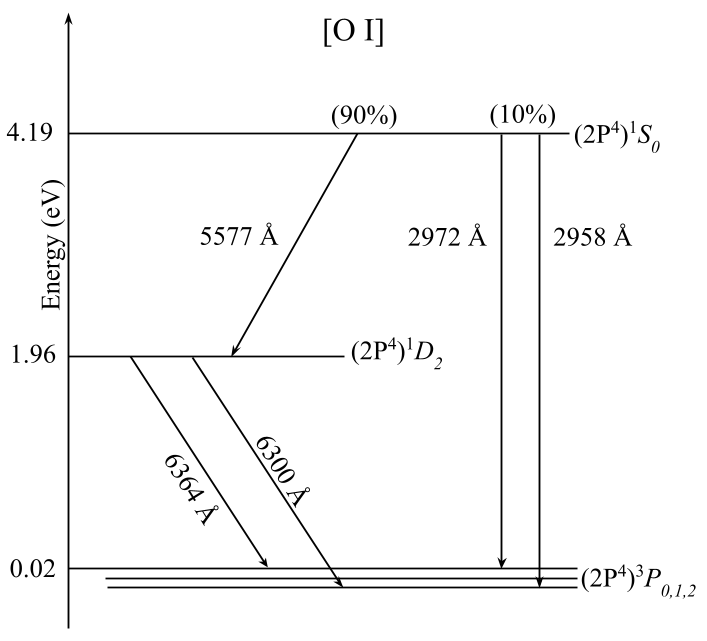}
    \caption{Energy level diagram for atomic oxygen [\ion{O}{1}]. In analysis of the oxygen line ratio, the 2972 and 2958\AA~lines can be ignored as they only comprise 10\% of transitions from the $^1S$ level to the $^1D$ level. Note that all atoms that decay through the 5577\AA~line will then also decay through either the 6300 or 6364\AA~line. Figure adapted from \cite{bhardwaj2012coupled}.}
    \label{int-fig:energylevels}
\end{figure}

Photodissociation of H$_2$O, CO$_2$, and CO causes the release of oxygen atoms in an excited state, either $^1S$ or $^1D$, depending on the wavelength of the incident photon. These atoms will then radiatively de-excite through the 5577\AA~line ($^1S$) or 6300 and 6364\AA~lines ($^1D$). Water releases oxygen atoms in the $^1S$ state at 3-8\% the rate that it releases atoms in the $^1D$ state, while CO$_2$ and CO release $O(^1S)$ at 30-90\% the rate of release of $O(^1D)$ \citep{delsemme1980photodissociation,festou1981forbidden,bhardwaj2012coupled}. This difference is reflected in the ratio of the line intensities (hereafter referred to as the "oxygen line ratio"), which is given by:
\begin{equation}
\label{int-eq:OIratio-initialdef}
    R \equiv \frac{N(O(^1S))}{N(O(^1D))} = \frac{I_{2958}+I_{2972}+I_{5577}}{I_{6300}+I_{6364}}
\end{equation}

\noindent
where $N(x)$ represents the column density of the species $x$ and $I_y$ represents the intensity of line $y$. Since the 2958 and 2972\AA~lines are much fainter than the others (10\% of the 5577\AA~line) and are unobservable from the ground, they are often ignored in calculations of the oxygen line ratio \citep{slanger2011atomic}, although one can also incorporate a correction factor for their absence into calculations of the $\frac{CO_2}{H_2O}$ ratio (see section \ref{meth-subsubsect:opt-analysis}). For low number densities where collisional quenching is not significant, the oxygen line ratio will always be less than one, because any atom that decays through the 5577\AA~line will then also decay through either the 6300 or 6364\AA~line (shown in Figure \ref{int-fig:energylevels}, the energy level diagram for [\ion{O}{1}]). If water is the dominant parent molecule, the oxygen line ratio calculated should be 0.03-0.08, whereas if  CO$_2$ or CO dominates, the oxygen line ratio should be 0.3-0.9 \citep{delsemme1980photodissociation,festou1981forbidden,bhardwaj2012coupled}.

This qualitative method of determining the dominant parent molecule has been well-established, but in recent years it has been suggested that the oxygen line ratio could also be used quantitatively to infer a relative abundance of CO$_2$ to H$_2$O \citep{mckay2012forbidden,mckay2013observations,mckay2015evolution,mckay2016co,decock2013forbidden}. However, this is only valid if the photochemistry governing [\ion{O}{1}] release is well understood. Despite attempts to develop and constrain [\ion{O}{1}] release rates in the laboratory, using these rates often produces discrepancies with direct measurements of CO$_2$ abundances from spacecraft, and empirical release rates determined in conjunction with these space-based measurements are often able to more accurately reproduce the directly measured $\frac{CO_2}{H_2O}$ ratio \citep{mckay2012forbidden,mckay2013observations,mckay2015evolution,mckay2016co}. However, empirical release rates only succeed at reproducing these direct measurements for some comets and not others. There was some success using the empirical release rates developed by \cite{mckay2015evolution} to reproduce direct measurements of comets C/2009 P1 (Garradd) and C/2012 K1 (PanSTARRS) \citep{mckay2015evolution,mckay2016co}. In more recent work, though, this same model did not successfully reproduce measurements of comets C/2013 US10 (Catalina) and C/2013 X1 (PanSTARRS) \citep{mckay2020establishing,hall2021using}.

One notable difference between these sets of observations was the solar cycle phase at the time of observation. Garradd and K1 PanSTARRS were observed during solar active periods, while Catalina and X1 PanSTARRS were observed during solar quiet periods. Therefore, it is possible that the solar cycle phase at the time of observation has some effect on release rate accuracy. The release of oxygen atoms into the coma is caused by photodissociation of the parent molecule by ultraviolet wavelengths of sunlight. The ultraviolet spectrum of the sun varies in spectral shape and intensity over the solar cycle. This variable behavior across a solar cycle could have an effect on the photochemistry of [\ion{O}{1}] release and may explain the discrepancies found between sets of release rates in the literature.

We present observations and analysis of the forbidden oxygen lines at 5577, 6300, and 6364\AA~in one comet observed near solar maximum, C/2014 Q2 (hereafter "Lovejoy"), and one near solar minimum, C/2007 N3 (hereafter "Lulin"). We use both the theoretical release rates from \cite{bhardwaj2012coupled} and the empirical release rates established by \cite{mckay2015evolution} to derive inferred CO$_2$ abundances for these comets and compare to contemporaneous space-based measurements of CO$_2$ obtained with the Spitzer Space Telescope (Lovejoy, this work) and AKARI spacecraft \citep[Lulin,][]{ootsubo2012akari}. We describe our observations and our data reduction and analysis procedures in Section \ref{sect:methods}. Section \ref{sect:results} presents our measured oxygen line ratios and a comparison of our inferred CO$_2$ abundances with the abundances measured by Spitzer and AKARI. Section \ref{sect:discussion} discusses the implications of our results, including the oxygen line ratio’s potential dependence on the solar cycle. In Section \ref{sect:conclusion} we summarize our conclusions and suggest future work.

\section{Observations and data analysis}
\label{sect:methods}

\subsection{Optical observations}
\label{meth-subsect:optical-obs}

We obtained spectra of both Lovejoy and Lulin across multiple dates. Table \ref{meth-tab:observation-table} provides the observation dates and geometries. The majority of the data were obtained using the Tull Coud\'e spectrograph mounted on the 2.7-meter Harlan J. Smith Telescope at McDonald Observatory. The Tull Coud\'e provides a spectral resolving power of $R \equiv \frac{\lambda}{\Delta \lambda}~ = 60000$ and a spectral range from 3500-10000\AA. Although this range contains interorder gaps redward of 5800\AA, we were careful to set the grating such that the red oxygen lines (6300 and 6364\AA) were fully covered by our observations. The size of the slit on the sky is 1.2" × 8.2". Other specifications concerning the Tull Coud\'e spectrograph can be found in \cite{cochran2001observations,tull1995high-resolution}. An additional night of observations for Lovejoy was obtained three months after the initial observations using the ARCES echelle spectrograph on the 3.5-meter Astrophysical Research Consortium Telescope at Apache Point Observatory. ARCES provides a spectral resolving power of R = $31500$ and a spectral range 3500-10000\AA. This range contains no interorder gaps. The projected size of the slit on the sky is 3.2" × 1.6". Other specifications concerning the ARCES spectrograph can be found in \citet{wang2003arces}, \citet{mckay2012forbidden}, and \citet{mckay2013observations}.

\begin{table*}
\caption{\label{meth-tab:observation-table}Optical observation log. For non-photometric nights, the spectrum of the flux standard star from either the previous or following night was used. However, because all lines are measured simultaneously, whether the night was photometric does not affect the ratio of the fluxes (i.e. the measured oxygen line ratios).}
\hspace*{-1cm}
\begin{tabular}{cccccccccc}
\hline
\multicolumn{1}{c}{Object} & \multicolumn{1}{c}{Date (UT)} & \multicolumn{1}{c}{$r$ (AU)} & \multicolumn{1}{c}{$\Delta$ (AU)} & \multicolumn{1}{c}{$\dot{\Delta}$ (km/s)} & \multicolumn{1}{c}{Instrument} & \multicolumn{1}{c}{Solar standard} & \multicolumn{1}{c}{Fast Rot.} & \multicolumn{1}{c}{Flux Cal.} & \multicolumn{1}{c}{Photometric?} \\
\hline
Lovejoy & 2015 Feb 02 & 1.29 & 0.79 & 32.9 & Tull Coud\'e & Solar port & HR 838 & \nodata & No \\
 & \multicolumn{1}{c}{2015 Feb 03$^a$} & 1.29 & 0.81 & 33.2 & Tull Coud\'e & Solar port & HR 628 & HR 3454 & Yes \\
 & 2015 Feb 04 & 1.29 & 0.83 & 33.4 & Tull Coud\'e & Solar port & \nodata & \nodata & No \\
 & 2015 May 11 & 1.95 & 2.16 & 12.7 & ARCES & HR 245 & HR 333 & HR 5501 & Yes \\
Lulin & 2009 Feb 14 & 1.32 & 0.56 & -43.4 & Tull Coud\'e & Solar port & \nodata & \nodata & No \\
 & 2009 Feb 15 & 1.33 & 0.54 & -41.1 & Tull Coud\'e & Solar port & HR 4875 & HR 4875 & Yes \\
\hline
\end{tabular} \\
\\
$^a$ Note: one of the spectra obtained on February 3 was obtained at an offset position 100" tailward; see Section \ref{sect:discussion} for analysis of this spectrum.
\end{table*}

We centered the slit on the optocenter of the comet for all observations, except for one spectrum of Lovejoy that was obtained at an offset position 100" tailward on February 3 (see Section \ref{sect:discussion} for discussion of this spectrum). We utilized an ephemeris from JPL Horizons to track the optocenter's apparent motion across the sky. The guiding software employs a boresight technique, which keeps the slit on the optocenter by monitoring optocenter flux that falls outside the slit. In order to correct for solar absorption and continuum features caused when comet dust reflects sunlight, a solar analog star (HD245) was observed for the ARCES data. Observing a solar standard star was not necessary for the Tull Coud\'e data, as the Tull Coud\'e spectrograph possesses a solar port that feeds in sunlight directly from the daytime sky, from which we could obtain an observed solar spectrum to correct for solar features. For most nights, we observed a fast-rotating ($vsin(i) > 150~km/s$) O, B, or A star to correct for telluric features in the spectra. If it was not possible to observe a fast rotator due to clouds or other hindrances, the fast rotator spectrum from the previous or following night was used when correcting for telluric features. For photometric nights, observations of a flux standard star were obtained, providing a means for absolute flux calibration for those nights. Table \ref{meth-tab:observation-table} lists the calibration stars observed and indicates whether conditions were photometric for each night. Spectra of a quartz lamp and ThAr lamp were also obtained for flat fielding and wavelength calibration, respectively.

\subsubsection{Data reduction}
\label{meth-subsubsect:opt-reduction}

We reduced and extracted the spectra using Image Reduction and Analysis Facility (IRAF) scripts that carry out bias subtraction, cosmic ray removal, flat fielding, and wavelength calibration procedures. To remove telluric absorption features, we divided the comet, solar, and flux standard spectra by the fast rotator spectrum. For photometric nights, we used the flux standard spectrum to convert the tellurically corrected comet spectrum flux to physical units. We shifted the tellurically corrected solar spectrum in wavelength and scaled it vertically to align with the flux-calibrated comet spectrum. We then subtracted this shifted, scaled solar spectrum from the comet spectrum to correct for solar absorption features and remove Fraunhofer lines and the continuum level. 

For spectrographs with narrow slit widths such as Tull Coud\'e and ARCES, it is necessary to estimate how much flux was lost falling outside of the slit during observations of the flux standard star. As a best practice, we incorporate a slit loss correction factor into analysis of all of our data. For observations of Lovejoy, we were able to find the transmittance through the slit by performing aperture photometry on the slit viewer images as described in \cite{mckay2014rotational}. Estimates of 75-90\% are typical for the transmittance, with a typical standard deviation in this estimate of around 10\%. Therefore, we adopt a 10\% uncertainty in our absolute flux calibration measurements (and therefore our water production rates). For observations of Lulin, slit viewer images were not available, so the slit loss correction was an estimate based on the average seeing, which was $\sim0.5$". 

\subsubsection{Deblending of telluric and cometary lines}
\label{meth-subsubsect:opt-deblending}

\begin{figure}
    \centering
    \includegraphics[width=\columnwidth]{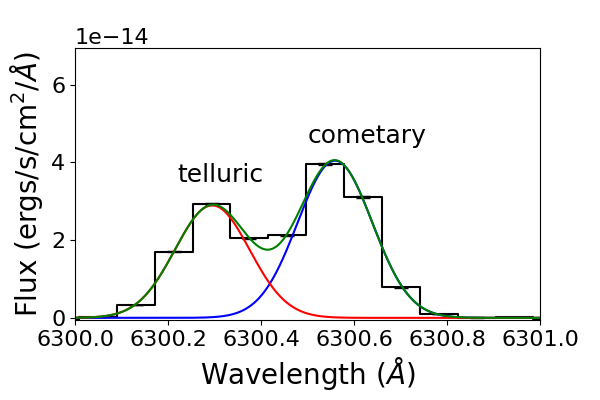}
    \caption{Region of the spectrum obtained of Lovejoy on UT May 11, 2015 that contains the forbidden oxygen line at 6300\AA. Note that the cometary line appears redward of 6300\AA~ due to Doppler shifting caused by the comet's geocentric velocity. The data are represented by a histogram (black) with $1\sigma$ errorbars. Due to the high signal-to-noise, the error bars are quite small, with typical values around 1\% of the flux. The data are fitted with two separate Gaussians corresponding to the telluric (red) and cometary (blue) features. The sum of the two Gaussian components is plotted in green.}
    \label{meth-fig:blended-lines}
\end{figure}

The forbidden oxygen lines at 5577, 6300, and 6364\AA~also occur as telluric emission features, meaning that a combination of a large geocentric velocity (and therefore a large Doppler shift) and high spectral resolution are necessary to separate the cometary and telluric lines. For almost all of the observations, the geocentric velocity was high enough that the cometary and telluric features were completely separated. However, for the May observations of Lovejoy, there was enough overlap between the lines that deblending them became necessary. Each line profile is determined by the instrumental point spread function because the line widths are unresolved. The instrumental point spread function is described well by a Gaussian for our spectra. We fit the observed line profile to the sum of two Gaussians (one corresponding to the cometary line and the other to the telluric line) using the Python package \textit{pyspeckit}’s function \textit{Spectrum}. A full description of the deblending process can be found in \cite{mckay2012forbidden}. Figure \ref{meth-fig:blended-lines} shows an example spectrum with this deblending method applied and the resulting curve fits constructed graphically.

\subsubsection{Data analysis}
\label{meth-subsubsect:opt-analysis}

The ratio of the 6300\AA~line flux to the 6364\AA~line flux reflects the branching ratio of the two respective transitions from the $^1D$ energy level to the $^3P$ ground state (see Figure \ref{int-fig:energylevels}). This ratio is well established as 3.0 \citep{storey2000theoretical,sharpee2006o1d2,cochran2001observations,cochran2008atomic,mckay2012forbidden,mckay2013observations,decock2013forbidden}. We confirmed that our data reproduced this accepted ratio within uncertainty before proceeding with further analysis. After this was confirmed, we used the measured line fluxes to calculate the oxygen line ratio, simplified from Equation \ref{int-eq:OIratio-initialdef} to be:
\begin{equation}
    R = \frac{I_{5577}}{I_{6300}+I_{6364}}
    \label{meth-eq:OIratio-simplified}
\end{equation} 

The ratio of CO$_2$ to H$_2$O can be derived by observing that each species’ contribution of line flux can be expressed as the product of the release rate $W$ and the column density $N$. Substituting these expressions for the line fluxes into Equation \ref{meth-eq:OIratio-simplified} and solving for $\frac{N_{CO_2}}{N_{H_2O}}$ produces the following equation \citep{mckay2012forbidden,decock2013forbidden}:
\begin{equation}
    \frac{N_{CO_2}}{N_{H_2O}} = \frac{RW^{red}_{H_2O} - W^{green}_{H_2O} - W^{green}_{CO}\frac{N_{CO}}{N_{H_2O}} + RW^{red}_{CO}\frac{N_{CO}}{N_{H_2O}}}{W^{green}_{CO_2}-RW^{red}_{CO_2}}
    \label{meth-eq:CO2-H2O-ratio-initial-def}
\end{equation}
\noindent
where $N_x$ represents the column density of species $x$ and $R$ represents the oxygen line ratio from Equation \ref{meth-eq:OIratio-simplified}. The release rate $W$ is defined as
\begin{equation}
    W = \tau^{-1}\alpha\beta
    \label{meth-eq:release-rate-def}
\end{equation}
\noindent
where $\tau$ is the parent molecule's photodissociative lifetime, $\alpha$ is the yield into the excited state of interest, and $\beta$ is a given line's branching ratio out of a certain excited state. Each instance of $W$ in Equation \ref{meth-eq:CO2-H2O-ratio-initial-def} specifies both the line in question (red or green, corresponding to the 6300/6364\AA~"red doublet" or the 5577\AA~"green line", respectively) and the parent molecule. We scale the release rate for $^1S$ (the green line) by 0.9 to account for the 10\% of atoms that de-excite through the unobservable 2972 and 2958\AA~lines. In this work we use both theoretical release rates from \cite{bhardwaj2012coupled} and empirical release rates from \cite{mckay2015evolution} (from which we use release rates "B", as this set seemed to prove more accurate when utilized by \cite{mckay2016co}) to infer CO$_2$ abundances from our oxygen line ratio measurements in order to evaluate the accuracy of both sets of release rates when used for objects observed at different solar cycle phases (see Table \ref{res-tab:release-rates} for numerical values of these release rates).

\begin{table}
\caption{\label{res-tab:release-rates}[\ion{O}{1}] release rates for H$_2$O and CO$_2$ for the $^1S$ and $^1D$ electron states. Column 3 lists empirical release rates (B) from \cite{mckay2015evolution}, while Column 4 lists theoretical release rates from \cite{bhardwaj2012coupled}.}
\begin{tabular}{cccc}
\hline
Parent & [\ion{O}{1}] State & McKay2015 & B\&R2012  \\
& & $(10^{-8}~s^{-1})$ & $(10^{-8}~s^{-1})$ \\
\hline
H$_2$O & $^1S$ & 0.64 & 2.6 \\
H$_2$O & $^1D$ & 84.4 & 84.4 \\
CO$_2$ & $^1S$ & 50.0 & 72.0 \\
CO$_2$ & $^1D$ & 75.0 & 120.0 \\
\hline
\end{tabular}
\end{table}

Although other, more complex molecules containing oxygen (e.g. H$_2$CO, CH$_3$OH) are present in cometary comae, they are much less abundant than H$_2$O, CO$_2$, and CO, and they release oxygen through a multi-step process. This makes contribution to the [\ion{O}{1}] population from these molecules very inefficient, and we consider this contribution negligible for our analysis. There is also some precedent for considering the contribution of CO photodissociation to the [\ion{O}{1}] population to be negligible \citep{raghuram2014photochemistry,Raghuram2020}. This is a reasonable assumption for our data due to the small measured $\frac{CO}{H_2O}$ ratios for Lovejoy and Lulin reported in other works: $(2.16~\pm~0.20)\%$ and $(2.17~\pm~0.11)\%$, respectively \citep{delloRusso2022volatile,gibb2012chemical}. Applying this assumption, Equation \ref{meth-eq:CO2-H2O-ratio-initial-def} simplifies to:
\begin{equation}
    \frac{N_{CO_2}}{N_{H_2O}} = \frac{RW^{red}_{H_2O} - W^{green}_{H_2O}}{W^{green}_{CO_2}-RW^{red}_{CO_2}}
    \label{meth-eq:CO2-H2O-ratio-simplified}
\end{equation}
\noindent
The relationships given in Equations \ref{meth-eq:CO2-H2O-ratio-initial-def} and \ref{meth-eq:CO2-H2O-ratio-simplified} are independent of heliocentric distance, meaning that successful application of this methodology does not depend on the heliocentric distance of the comet \citep{mckay2013observations,mckay2015evolution,mckay2016co}. 

For small fields of view (which applies to both our ARCES and Tull Coud\'e data), the column density ratio from Equation \ref{meth-eq:CO2-H2O-ratio-simplified} reflects the production rate ratio (see \cite{mckay2015evolution} and references therein for more details). A small field of view can also increase the importance of accounting for the preferential collisional quenching of $^1D$ atoms (corresponding to the 6300 and 6364\AA~lines) over $^1S$ atoms (corresponding to the 5577\AA~line) \citep{bhardwaj2012coupled,raghuram2014photochemistry,decock2014forbidden}. The oxygen line ratio used in the previous equations are calculated under the assumption that collisional quenching did not affect the line fluxes, which is likely untrue. Thus, the observed fluxes for the 6300 and 6364\AA~lines need to be increased to account for the $^1D$ atoms that were present in the coma but de-excited through collisions instead of emitting photons (and therefore do not contribute to the intensities of the observed emission lines). The dominant molecule contributing to collisions is H$_2$O, so to estimate the amount of collisional quenching in an observation, we must estimate the H$_2$O production rate.

H$_2$O production rates were determined with our 6300\AA~observations using the Haser model \citep{haser1957distribution} and other established algorithms described in \cite{mckay2012forbidden,mckay2015evolution}. We adopt an expansion velocity following the relation from \cite{cochran1993observational,budzien1994solar}:
\begin{equation}
    \label{meth-eq:haser-model}
    v=(0.85~\mathrm{km~s^{-1}})\,r^{-0.5}
\end{equation}
\noindent
where $r$ is the heliocentric distance of the comet in units of AU. The algorithms employed produce a correction factor that scales the observed flux to the expected theoretical flux without collisional quenching. This correction factor is then applied to our oxygen line ratio measurements. However, while our oxygen line ratios are corrected for collisional quenching, the Haser model itself does not explicitly account for effects of collisions on the expansion velocity. This has the potential to result in overestimates of water production, which we discuss further in Section 4.

To evaluate any effects on our derived water production rates introduced by the uncertainties in [\ion{O}{1}] photochemistry, we compared our results with other measurements of water production rates for Lovejoy and Lulin obtained around the same dates as our own observations (see Section \ref{sect:discussion}). Additionally, using the 6300\AA~ line flux has been well established as a proxy for obtaining water production rates \citep{morgenthaler2001large,morgenthaler2007large,fink2009taxonomic,mckay2015evolution,mckay2018evolution,mckay2021quantifying}. The original and corrected oxygen line ratios are presented in Section \ref{sect:results} along with our water production rates.

\subsection{Infrared observations}
\label{meth-subsect:infrared-obs}
In addition to the optical observations obtained for oxygen line ratio measurements, infrared observations of comet Lovejoy were obtained in order to provide a direct measurement of CO$_2$ for comparison. These observations were obtained with the Spitzer Space Telescope \citep{werner04-spitzer} as part of the Cometary Orbital Trends with Spitzer survey \citep[program IDs 11106 and 13116;][]{kelley17-dps}. Images were obtained with the InfraRed Array Camera (IRAC) instrument, which is comprised of four 256$\times$256 pixel arrays (1\farcs22 per pixel) in the focal plane, each with their own broadband filter \citep{fazio04-irac}. Observations were obtained after Spitzer's cryogens were depleted, and were limited to two bandpasses: 3.6~\micron{} and 4.5~\micron{}. The comet was observed on three dates near perihelion, which occurred on 2015 January 30 at 01:40 UTC (JPL Horizons solution \#59). The observational circumstances are presented in Table~\ref{meth-tab:infrared-obs}. The telescope followed the comet with its non-sidereal rates, and mapped the coma with a 3$\times$3 array pattern with 260\arcsec{} steps (84\% of the array size). The two detector arrays observe independent locations on the sky separated by 91\arcsec{}; therefore, the pattern was offset in such a way to enable efficient mapping of the comet with a single observing sequence. At each position in the pattern, 5 images were taken, each with random offsets to reduce noise and improve scene sampling \cite[IRAC's ``cycling'' dither pattern with a medium scale;][]{irac}. Two exposures were taken at each position with 0.4 and 10.4~s exposure times, but the longer exposures saturated at 4.5~\micron{} so we limited our analysis to the short exposure data. To measure the background, the observing sequences were re-executed at the same location on the sky, including non-sidereal rates, after the comet had moved out of the mapped area.

\begin{table*}
\caption{\label{meth-tab:infrared-obs} Observational circumstances and photometric data for the Spitzer Space Telescope and AKARI data sets.}
\hspace*{-2.5cm}
\begin{tabular}{cccccccccccc}
\hline
Comet & Telescope & Date & $r$ & $\Delta$ & $F_\nu(3.6)^a$ & $F_\nu(4.5)^a$ & $C_{\rm dust}$$^b$ & $Q({\rm CO_2})$ & Reference \\
& & (UTC) & (AU) & (AU) & (Jy) & (Jy) &  & ($10^{28}$ s$^{-1}$) & \\
\hline
Lulin   & AKARI   & 2009 Feb 05 00:34 & 1.275 & 0.817 & \nodata & \nodata & \nodata & $0.485\pm0.049$ & Ootsubo et al. 2012 \\
Lovejoy & Spitzer & 2015 Feb 17 03:28 & 1.319 & 0.974 & $2.37\pm0.07$ & $9.99\pm0.30$ & 2.84 & $4.95\pm0.15$ & This work \\
        & Spitzer & 2015 Feb 27 15:19 & 1.360 & 1.002 & $2.21\pm0.07$ & $9.02\pm0.27$ & 2.65 & $5.15\pm0.15$ & \\
        & Spitzer & 2015 Mar 13 04:08 & 1.436 & 1.130 & $1.59\pm0.05$ & $6.41\pm0.19$ & 2.24 & $5.69\pm0.17$ & \\
\hline
\end{tabular}\\
\noindent$^a$Not color corrected.  $^b$Dust 3.6-to-4.5~\micron{} scale factor.
\end{table*}

\begin{figure*}
    \centering
    \includegraphics[width=0.8\linewidth]{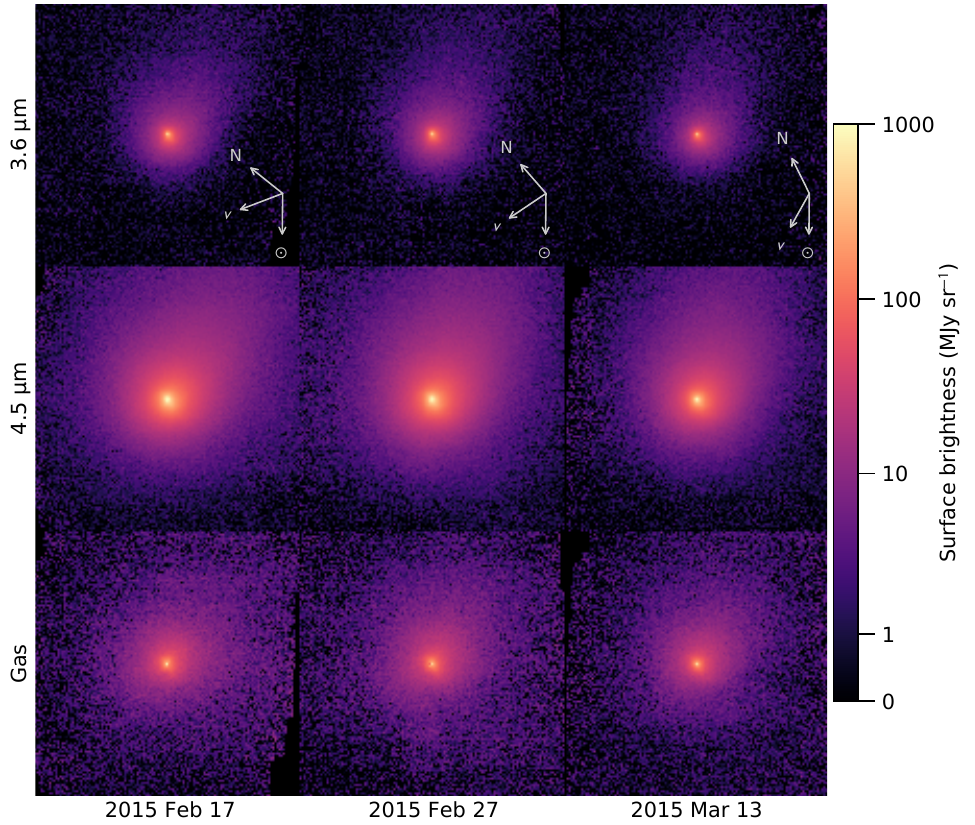}
    \caption{Spitzer IRAC images of comet C/2014 Q2 (Lovejoy). Top row: 3.6~\micron{} images. Center row: 4.5~\micron{} images. Bottom row: derived CO$_2$ gas maps. All images are displayed with an asinh color scale intended to show both the low-surface brightness instrumental background and the bright cometary core. Arrows indicate the projected Celestial north (N), Sun ($\sun$) and velocity ($v$) vectors. Each sub-panel is 8\farcm1$\times$8\farcm1.}
    \label{meth-fig:spitzer}
\end{figure*}

Raw images were processed with version S19.2.0 of the IRAC pipeline to produce flux calibrated images and downloaded from the Spitzer Heritage Archive \citep{spitzer-heritage-archive}. We use the pipeline's ``corrected'' data products, which have been produced to address several common artifacts: stray light, multiplexer striping, column pull down, and banding (internal optical scattering). The individual images were combined in the rest frame of the comet by date, exposure time, and bandpass using the MOPEX software \citep{makovoz05-mopex}. After subtracting the shadow mosaics from the target mosaics, residual background was measured in an area clear of the comet and removed. The resulting images of the comet are presented in Fig.~\ref{meth-fig:spitzer}.

IRAC's broadband filters may include significant contributions from dust (scattered light and/or thermal emission), and gas, especially CO$_2$ at 4.26~\micron{} and CO at 4.67~\micron{} \citep{reach13-coco2}. The gas can be isolated by scaling the 3.6~\micron{} image to match the dust brightness in the 4.5~\micron{} image, then subtracting the two. Following previous work, we use a morphological technique to determine the scale factor, assuming the dust morphologies of the two images are the same \citep{mckay2016co}. The gas coma is accounted for with a Haser model using the photodissociation rates for CO$_2$ from \citet{huebner1992solar}. The photometric aperture size used for analysis of these observations is much larger than the collisional radius, so we assume collisional equilibration, and thus for this Haser model we use the same expansion speed used for the water. We discuss potential caveats of this choice in Section \ref{sect:discussion}. Active sun photodissociation rates ($4.76\times10^{-6}$~s$^{-1}$ at 1 AU) produced a better fit to the coma than quiet sun rates, and were ultimately used for this work. The images are divided into a cylindrical grid with 30 radial steps (from 10 to 100 pixels) and 32 azimuthal steps (from 0 to 360\degr). For each azimuthal step, the radial profiles of the 3.6~\micron{} image, 4.5~\micron{} image, and the model gas coma are calculated. With a least squares optimization technique, the scale factors for the dust and each azimuth bin of the gas coma are computed. A gas-only image is made and the coma brightness measured in a 24\farcs4 (20 pix) radius aperture. The brightness is converted to production rate with a fluorescence $g$-factor of $2.69\times10^{-3}$ photons~s$^{-1}$ at 1 AU from the Sun \citep{debout16}, and the same Haser model as the morphological analysis. Uncertainties are based on IRAC pipeline uncertainty images. An additional 3\% uncertainty is assumed based on the absolute calibration of the instrument \citep{reach05-irac}. Photometry, dust scale factors, and production rates are given in Table~\ref{meth-tab:infrared-obs}. The photometry is not color corrected in order to preserve the CO$_2$ irradiance.

For comet Lulin, a direct CO$_2$ measurement for comparison to our inferred value was made by \cite{ootsubo2012akari}. Their observations were obtained in the near-infrared using AKARI. The measurement of theirs that we reference in this work was obtained on UT 2009 February 5, around ten days before our own measurements. Observation geometries are provided in Table \ref{meth-tab:infrared-obs}, with more details in \cite{ootsubo2012akari}.

\section{Results}
\label{sect:results}

\begin{figure*}
\includegraphics[width=2\columnwidth]{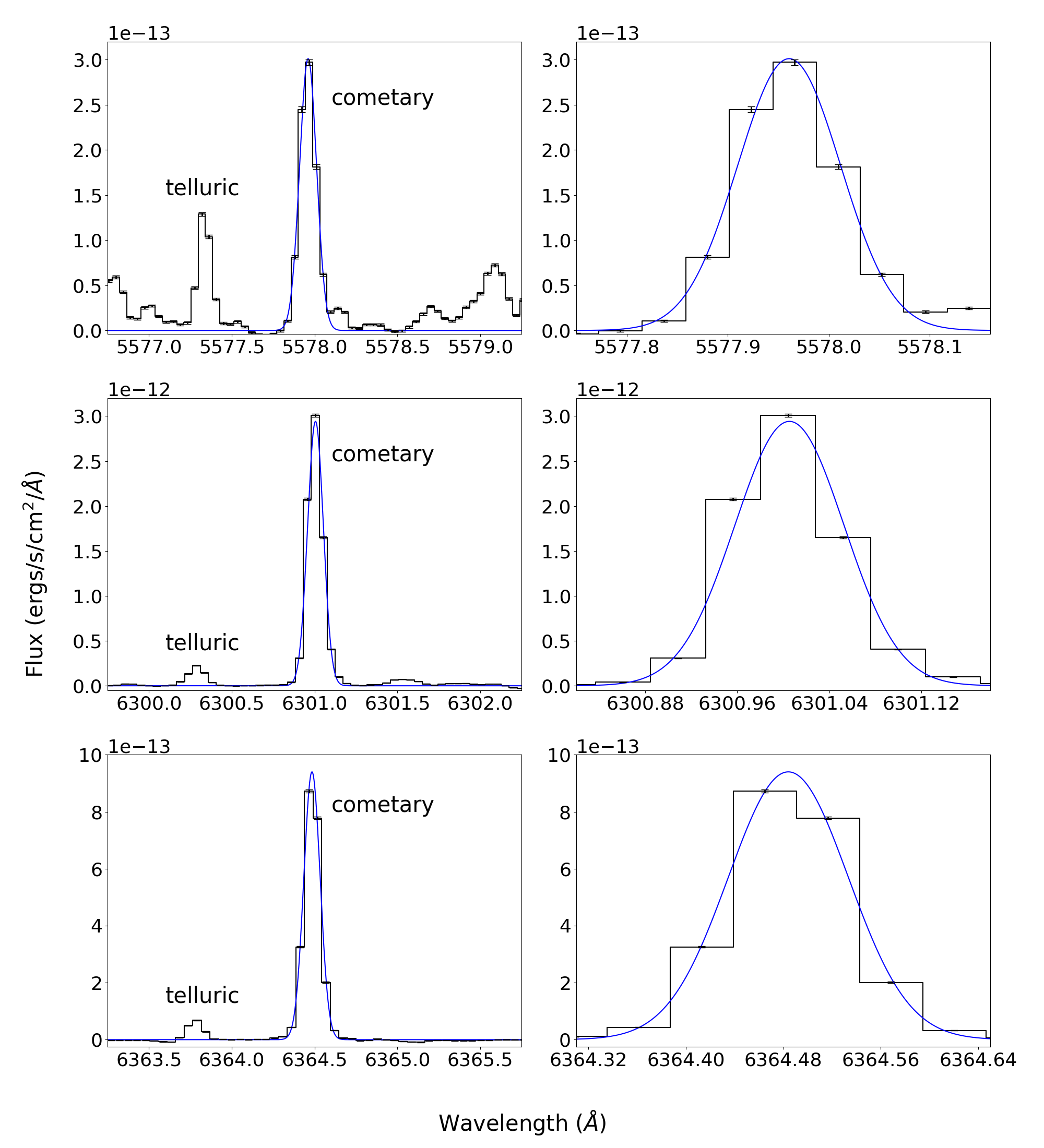}
\caption{\label{res-fig:prettyfit-doublepanel-lovejoy}Regions of the spectra obtained of Lovejoy on UT 2015 February 3 that contain the forbidden oxygen lines at 5577\AA~(top row), 6300\AA~(middle row), and 6364\AA~(bottom row). Note that the cometary line appears redward of the telluric in all plots due to Doppler shifting caused by the comet's geocentric velocity. The data are represented by a histogram (black) with $1\sigma$ errorbars. The right column shows the cometary lines fitted with Gaussian curves (blue). The left column displays a zoomed-out view such that both the cometary and telluric lines are visible as well as the surrounding signal-to-noise. In the top left panel, the weaker features surrounding the two oxygen lines are due to emission from C$_2$.}
\end{figure*}

\begin{figure*}
\includegraphics[width=2\columnwidth]{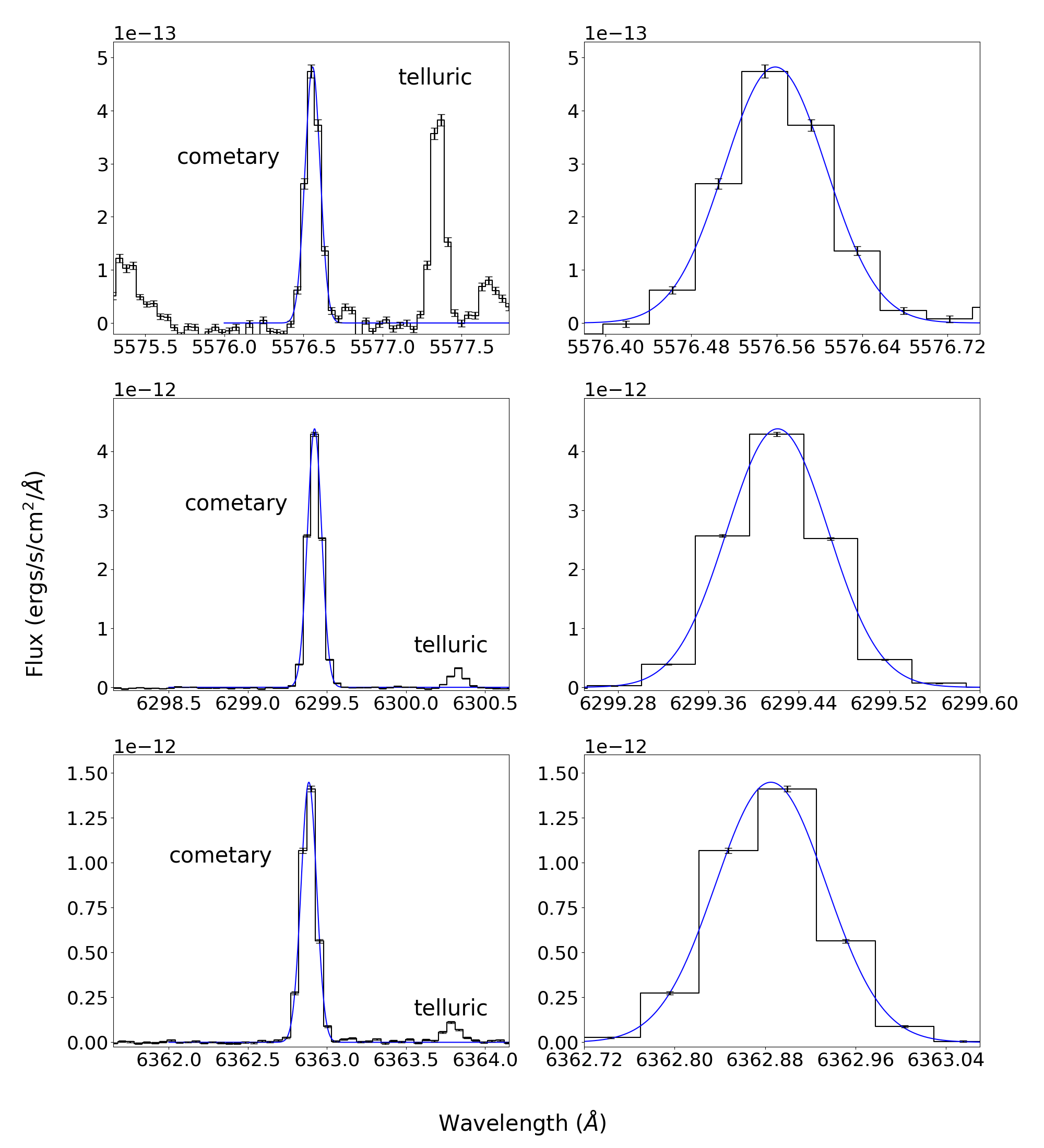}
\caption{\label{res-fig:prettyfit-doublepanel-lulin}Regions of the spectra obtained of Lulin on UT 2009 February 14 that contain the forbidden oxygen lines at 5577\AA~(top row), 6300\AA~(middle row), and 6364\AA~(bottom row). Note that the cometary line appears blueward of the telluric in all plots due to Doppler shifting caused by the comet's geocentric velocity. The data are represented in the same ways described in Figure \ref{res-fig:prettyfit-doublepanel-lovejoy}. In the top left panel, the weaker features surrounding the two oxygen lines are due to emission from C$_2$.}
\end{figure*}

We display a subset of our spectra for Lovejoy and Lulin with lines of best fit overplotted in Figures \ref{res-fig:prettyfit-doublepanel-lovejoy} and \ref{res-fig:prettyfit-doublepanel-lulin}. We present initial oxygen line ratio measurements across multiple dates for each comet in Table \ref{res-tab:OI-ratios}. For photometric nights, we report H$_2$O production rates, from which we calculated a collisional quenching (CQ) factor. We divided the observed oxygen line ratio by the collisional quenching factor to obtain a new value for the oxygen line ratio that has been adjusted to account for collisional quenching. Table \ref{res-tab:OI-ratios} shows our H$_2$O production rates, collisional quenching factors, and adjusted oxygen line ratios. 


For non-photometric nights, absolute flux calibration was not possible, and therefore direct calculation of an H$_2$O production rate and CQ factor was not possible. For these nights, we adopted the CQ factor from the previous or following photometric night in order to calculate an adjusted oxygen line ratio. CQ factors were not adopted from nights more than a day apart to minimize any change in the comet's true H$_2$O production rate. For the spectrum obtained of Lovejoy on February 3 at an offset position (hereafter the "offset spectrum"), we elected not to calculate an H$_2$O production rate or CQ factor. For offset spectra, the production rate derived using the Haser model would not necessarily reflect the global production rate, and it is not required for our analysis. We discuss this spectrum more generally in terms of its unmodified oxygen line ratio in Section \ref{disc-subsect:helio-cometo-centric-distance-changes}.

\begin{table*}
\caption{\label{res-tab:OI-ratios} Oxygen line ratios and H$_2$O production rates. The sixth column contains the collisional quenching factor, which is derived from the H$_2$O production rate and is used to correct the [\ion{O}{1}] ratio. For non-photometric nights, calculating the H$_2$O production rate is not reliable, so instead the CQ factor from the previous or following night was used to get the adjusted oxygen line ratio.}
\hspace*{1cm}
\begin{tabular}{ccccccc}
\hline
\multicolumn{1}{c}{Object} & \multicolumn{1}{c}{Date} & \multicolumn{1}{c}{$r$} & \multicolumn{1}{c}{[\ion{O}{1}] line ratio} & \multicolumn{1}{c}{H$_2$O prod. rate} & \multicolumn{1}{c}{CQ factor} & \multicolumn{1}{c}{Adjusted [\ion{O}{1}] ratio} \\
 & \multicolumn{1}{c}{(UT)} & \multicolumn{1}{c}{(AU)} & & \multicolumn{1}{c}{($10^{29} s^{-1}$)} & & \\
\hline
Lovejoy & 2015 Feb 02 & 1.29 & 0.089 $\pm$ 0.006 & \nodata & \nodata & 0.033 $\pm$ 0.003 \\
 & 2015 Feb 03 & 1.29 & 0.079 $\pm$ 0.005 & 6.57 $\pm$ 0.74 & 2.70 $\pm$ 0.10 & 0.029 $\pm$ 0.003 \\
 & \multicolumn{1}{c}{2015 Feb 03$^a$} & 1.29 & 0.175 $\pm$ 0.031 & \nodata & \nodata & \nodata \\
 & 2015 Feb 04 & 1.29 & 0.085 $\pm$ 0.006 & \nodata & \nodata & 0.031 $\pm$ 0.003 \\
 & 2015 May 11 & 1.95 & 0.065 $\pm$ 0.004 & 1.19 $\pm$ 0.11 & 1.36 $\pm$ 0.02 & 0.048 $\pm$ 0.004 \\
Lulin & 2009 Feb 14 & 1.32 & 0.079 $\pm$ 0.004 & \nodata & \nodata & 0.041 $\pm$ 0.003 \\ 
 & 2009 Feb 15 & 1.33 & 0.080 $\pm$ 0.003 & 1.09 $\pm$ 0.11 & 1.94 $\pm$ 0.06 & 0.041 $\pm$ 0.002 \\
\hline
\end{tabular} \\
\\
$^a$ Results for the "offset spectrum" obtained 100" tailward of the optocenter. We were not able to calculate a water production rate (and therefore a CO$_2$ abundance) for this spectrum because the Haser model only applies for water production at the center of the comet. \\ 
\end{table*}

We calculate average oxygen line ratios of $0.031~\pm~0.002$ for the February 2-4 observations of Lovejoy and $0.041~\pm~0.002$ for the February 14-15 observations of Lulin. We use our calculated oxygen line ratios to infer CO$_2$ abundances via Equation \ref{meth-eq:CO2-H2O-ratio-simplified} using the release rates from \cite{mckay2015evolution} and \cite{bhardwaj2012coupled} listed in Table \ref{res-tab:release-rates}. Table \ref{res-tab:CO2-abundances} shows CO$_2$ abundances for each comet inferred using the empirical release rates \citep{mckay2015evolution} and the theoretical release rates \citep{bhardwaj2012coupled}. In the final column, direct CO$_2$ measurements are shown from Spitzer (Lovejoy, this work) and AKARI \citep[Lulin,][]{ootsubo2012akari}, for comparison to the inferred abundances. For Lovejoy, we derive a direct CO$_2$ production rate based on a linear fit to the measured values extrapolated to February 3: $(4.47\pm0.25)\times10^{28}$~s$^{-1}$. We divide our direct CO$_2$ production rate measurement from Spitzer by our calculated H$_2$O production rate from the [\ion{O}{1}] 6300\AA~ line flux to get the "direct" measurement of the $\frac{CO_2}{H_2O}$ ratio, since H$_2$O was not observed with Spitzer.

\begin{table*}
\centering
\caption{\label{res-tab:CO2-abundances} CO$_2$ abundances inferred using the adjusted oxygen line ratios from Table \ref{res-tab:OI-ratios}. The abbreviation "emp." refers to the empirical release rates from \cite{mckay2015evolution}, and "th." refers to the theoretical release rates from \cite{bhardwaj2012coupled}. The two inferred CO$_2$ abundances are listed next to the directly measured CO$_2$ abundances ("dir.") from Spitzer (this work) and AKARI \citep{ootsubo2012akari}.}
\begin{tabular}{cccc|c}
\hline
\multicolumn{1}{c}{Object} & \multicolumn{1}{c}{Date} & \multicolumn{1}{c}{$\frac{CO_2}{H_2O}$ emp.} & \multicolumn{1}{c}{$\frac{CO_2}{H_2O}$ th.} & \multicolumn{1}{c}{$\frac{CO_2}{H_2O}$ dir.} \\
 & \multicolumn{1}{c}{(UT)} & \multicolumn{1}{c}{(\%)} & \multicolumn{1}{c}{(\%)} & \multicolumn{1}{c}{(\%)} \\
\hline
Lovejoy & 2015 Feb 2-4 & 7.2 $\pm$ 0.6 & 0.5 $\pm$ 0.3 & 7.4 $\pm$ 0.4$^a$ \\
 & 2015 May 11 & 12.7 $\pm$ 1.3 & 3.0 $\pm$ 0.6 & \nodata \\
Lulin & 2009 Feb 14-15 & 10.4 $\pm$ 0.7 & 1.9 $\pm$ 0.3 & 11.9 $\pm$ 1.7$^b$ \\
\hline
\end{tabular} \\
\vspace{0.1cm}
\raggedright
$^a$Abundance calculated using inferred CO$_2$ production rate for February 3 extrapolated from line fit to direct measurements obtained by Spitzer over three dates in 2015 February-March (see Section \ref{meth-subsect:infrared-obs}). \\
$^b$ Direct measurement obtained by AKARI on UT 2009 February 5. Cited from \cite{ootsubo2012akari}.
\end{table*}

\section{Discussion}
\label{sect:discussion}

\subsection{Lovejoy: Changes in [\ion{O}{1}] ratio with heliocentric and cometocentric distance}
\label{disc-subsect:helio-cometo-centric-distance-changes}

Our average measured oxygen line ratio for Lovejoy in May is about 40\% larger than the measured ratio in February, indicating that the oxygen line ratio increases with increasing heliocentric distance. This supports the well-established observation that H$_2$O sublimation falls off faster than CO$_2$ with heliocentric distance due to the higher volatility (lower sublimation temperature) of CO$_2$ \citep{meech2004using,ootsubo2012akari}. The increase in Lovejoy's oxygen line ratio from February to May is indicative of a similar trend for this comet. Comparing the CO$_2$ abundance inferred from the February data to the abundance inferred in May, the abundance increased by about 50\% using the empirical release rates and increased by almost 150\% using the theoretical release rates. These results show that CO$_2$ may start to become a more important driver of activity even at heliocentric distances as small as 2 AU.

For the February 3 Lovejoy data, the "offset spectrum" produced a raw (i.e., unadjusted for collisional quenching) oxygen line ratio of $0.175~\pm~0.031$, which is about a factor of 2 larger than the average raw oxygen line ratio of the other spectra obtained that same date that were centered on the comet's optocenter, which was $0.079~\pm~0.002$. Even though collisional quenching is not accounted for yet in the raw oxygen line ratios, we can assert that the higher ratio for the offset spectrum is not due to any collisional quenching effects. This is because the amount of collisional quenching that occurs scales with density, and any offset position in the coma will be much less dense than the optocenter, meaning much less collisional quenching will occur at the offset position. The increased oxygen line ratio also cannot be due to the presence of icy grains. The effect of an icy grain source would be to make the oxygen line ratio lower, not higher, since the icy grain source would create increased water production in the outer coma and therefore work to lower the observed oxygen line ratio. The increased oxygen line ratio could be due to an anti-sunward asymmetry, but because the species measured are neutral, one would not necessarily expect them to be enhanced along the tail. That being said, we cannot definitively rule out the possibility that an anti-sunward jet is responsible for our higher measured oxygen line ratio at the offset coma position. 



However, an increasing [\ion{O}{1}] ratio with increasing cometocentric distance follows logically from what is known about the scale lengths for photodissociation of H$_2$O vs. CO$_2$. In cometary comae, H$_2$O has a scale length of about 80,000 km, while CO$_2$ has a scale length of about 500,000 km due to the longer lifetime of CO$_2$ in the coma \citep{bockelee1989nature,huebner1992solar}. The spectrum's offset distance of 100" is equivalent to a linear distance of roughly 60,000 km, which is comparable to the H$_2$O scale length, but an order of magnitude less than the CO$_2$ scale length. It is reasonable to expect that with increasing distance from the center of the comet, H$_2$O will be lost preferentially over CO$_2$, and thus CO$_2$ will play a larger role in supplying the [\ion{O}{1}] population in the outer coma, increasing the oxygen line ratio for offset spectra such as the one observed in this work. This argument assumes that the expansion velocities are the same for both species, and if this is invalid, the increase we observe may not be due to this reasoning. However, unless the expansion velocities are drastically different (comparable to the difference in scale lengths, so roughly a factor of 10), which is unlikely, then we would still expect to see an oxygen line ratio increase as is observed here. Indeed, some analysis of this phenomenon has been conducted in other works \citep[e.g.,][]{nelson2021evolving}. However, more study is still needed to confirm this potential trend of increasing [\ion{O}{1}] ratio with cometocentric distance. In future work, more spatially resolved observations of the oxygen coma could shed more light on the dependence of the oxygen line ratio on cometocentric distance. 


\subsection{Water production rates}
\label{disc-subsect:water-production-rates}

\subsubsection{Lovejoy}
\label{disc-subsubsect:lovejoy-h2o-prodrates}

We now compare our calculated H$_2$O production rates for Lovejoy on 2015 February 3 to other published values for the same time period. \cite{biver2015ethyl} took observations of water with IRAM from January 30 to February 3 and reported an H$_2$O production rate of $(7.5~\pm~0.3)\times10^{29}~s^{-1}$. SOHO/SWAN observations on February 3 from \cite{combi2018water} resulted in a production rate of $(5.9140~\pm~0.6805)\times10^{29}~s^{-1}$. NIRSPEC observations on February 3 from \cite{delloRusso2022volatile} resulted in a production rate of $\sim(5$\textendash $8)\times10^{29}~s^{-1}$. NIRSPEC observations on February 4 and 5 by \cite{paganini2017ground} resulted in a production rate of $(5.9~\pm~0.13)\times10^{29}~s^{-1}$. All of these measurements agree with our own production rate of $(6.57~\pm~0.74)\times10^{29}~s^{-1}$ within uncertainty.

Other measurements from the literature do not agree with our water production value as closely. Near-infrared observations obtained by \cite{faggi2016detailed} from January 30 - February 2 produced a production rate of $(4.88~\pm~0.22)\times10^{29}~s^{-1}$. \cite{feldman2018far} reported a production rate of $(4.5~\pm~0.9)\times10^{29}~s^{-1}$ from HST observations obtained on February 2. \cite{biver2016isotopic} reported a production rate of $(8.7~\pm~0.7)\times10^{29}~s^{-1}$ using the \textit{Odin} submillimeter space observatory from January 29 - February 3. Despite the larger discrepancies, these measurements are all within $3\sigma$ of our derived value. The discrepancies are reduced further after considering additional uncertainty from possible rotational variation. The \textit{Odin} observations indicated significant variability in the water production rate across the observation date range, which overlaps very closely with our own range. They showed that the variability of the water production rate over this date range fit approximately to a sine wave with an amplitude of $\pm20\%$. Taking this variation into account as additional uncertainty, all three discrepant values above agree with our measurement.

We also compare our calculated H$_2$O production rate for Lovejoy on May 11 to an additional production rate measurement from \cite{combi2018water} on May 11. They reported an H$_2$O production rate of $(2.1820~\pm~0.0087)\times10^{29}~s^{-1}$, which is about a factor of two larger than our measurement of $(1.19~\pm~0.11)\times10^{29}~s^{-1}$. However, some previous works \citep[e.g.,][]{mckay2015evolution} have found SWAN to give systematically high water production rates, likely because it detects production from icy grains in the outer coma, which would not be included in our calculations of production in the inner coma. The agreement between our February values could be related to the shorter sublimation lifetimes of the icy grains at smaller heliocentric distances, which could result in the grains mostly sublimating within our observation aperture, eliminating the discrepancy between our smaller field observations and the wider field of view of SWAN. However, it is important to note that while these effects could explain the discrepancy in our water production rates, this would not cause any change in the [\ion{O}{1}] ratio or inferred CO$_2$ abundance, for the purposes of this work.

\subsubsection{Lulin}
\label{disc-subsubsect:lulin-h2o-prodrates}

We compare our calculated H$_2$O production rate for Lulin on 2009 February 15 to other works that have reported production rates for Lulin near this date. NIRSPEC observations from January 30 - February 1 by \cite{gibb2012chemical} reported production rates between $(1.13~\pm~0.02)\times10^{29}~s^{-1}$ and $(2.68~\pm~0.11)\times10^{29}~s^{-1}$, which agrees with our measurement of $(1.09~\pm~0.11)\times10^{29}~s^{-1}$ within uncertainty. However, other works reported production rates that do not agree with our measurement as closely. \textit{Swift} observations obtained by \cite{bodewits2010swift} on January 28 yielded production rates between $\sim(6.7$\textendash $7.9~\pm~0.7)\times10^{28}~s^{-1}$. AKARI observations from February 5 by \cite{ootsubo2012akari} resulted in a production rate of $(4.091~\pm~0.409)\times10^{28}~s^{-1}$. Narrowband photometry obtained on February 26 by \cite{bair2018coma} resulted in a production rate of $(6.0~\pm~0.3)\times10^{28}~s^{-1}$. These are each about a factor of 2 lower than our measurement.

\begin{figure*}
    \centering
    \includegraphics[width=2\columnwidth]{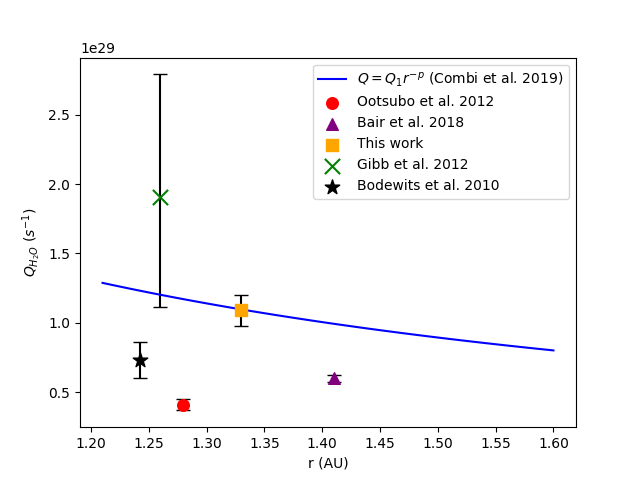}
    \caption{\label{disc-fig:lulin-h2o-fitting}Water production rates for Lulin from the literature, with the power law prediction by \cite{combi2019survey} for the relationship between water production rate and heliocentric distance post-perihelion overplotted. Some scatter is present in the data, with three measurements falling under the predictive curve from \cite{combi2019survey}. The uncertainty on the production rate from \cite{bair2018coma} represents the standard deviation of their measurements; however, the error bars are smaller than the size of the marker on the graph.}
\end{figure*}

Water production in Lulin around these dates is clearly complex, given these several discrepant values found by various sources. \cite{gibb2012chemical} found that their measured water production rate fluctuated significantly even over a timescale of hours, and began steadily increasing the last date of their observations (February 1). They suggested that this variability could be due to a rotating jet morphology. Additionally, \cite{combi2019survey} predicted a water production rate that decreases with heliocentric distance post-perihelion according to the power law:

\begin{equation}
    \label{disc-eq:waterprod-relation}
    Q = Q_1r^{-p}
\end{equation}

\noindent where $Q$ represents the water production rate, $Q_1$ represents the production rate at 1 AU, $r$ represents heliocentric distance, and $p$ represents the power-law exponent. For measurements of Lulin between 2009 January 10 and 2009 April 21, \cite{combi2019survey} reported $Q_1=1.78\times10^{29}$ and $p=-1.7\pm0.1$. We plot this relation over each of the water production rates for Lulin discussed here in Figure \ref{disc-fig:lulin-h2o-fitting}. There appears to be some scatter among the measurements, which could be a manifestation of variations in production rate with rotational phase. However, if the power law from \cite{combi2019survey} is assumed to be correct, there could be other factors at play causing the underestimates from some of the measurements. For example, an underestimate for water production measured from infrared observations (such as those reported in \cite{ootsubo2012akari}) could be due to optical depth effects that were unaccounted for, which may become significant at production rates on the order of $10^{29}~s^{-1}$ such as these. If the production rate from \cite{ootsubo2012akari} was indeed an underestimate due to optical depth effects, their calculated CO$_2$ abundance would have also been affected. Because we use this CO$_2$ abundance for comparison to our results, optical depth effects on their observations could complicate our interpretation. We discuss this further in Section \ref{disc-subsect:CO2-abundances}.

\subsubsection{Potential Caveats}
\label{disc-subsubsect:h2o-potential-caveats}

When comparing to other H$_2$O production rates in the literature, it is important to note a few methodology differences that could affect the comparison. First, we note that many of the other observations we compare to also calculated H$_2$O production rates using indirect methods, such as the measurement of OH or H. Many of these methods are model-dependent, and when different models are used, it adds scatter to the production rate values that would not necessarily be covered by stochastic error bars.

The presence of icy grain sublimation could also cause systematic differences between H$_2$O production rates measured using different methods. In the presence of icy grains, it has been shown that the H$_2$O production rate tends to correlate with projected size of the region of the coma observed \citep{mckay2015evolution}. For both Lovejoy and Lulin, the H$_2$O production rates we compare with cover a range of aperture sizes without demonstrating any systematic trend with aperture size. Thus, we conclude that any icy grain presence does not significantly affect our H$_2$O production rates compared with those obtained by other methods.

Lastly, calculating H$_2$O production rates from the 6300\AA~line flux is shown to be a generally reliable method in other works \citep{morgenthaler2001large,morgenthaler2007large,fink2009taxonomic,mckay2018evolution,mckay2021quantifying}. However, it is important to acknowledge the caveats of the method's use of the Haser model, particularly when analyzing comets with high production rates such as Lovejoy and Lulin. Collisions make the spatial distribution of molecules more centrally peaked, and thus by using the Haser model, our production rates could be overestimates of the true values. Furthermore, if our estimate of collisional equilibration between water and CO$_2$ is incorrect, this could have additional effects on the derived production rates. While these factors could be at play, our analysis suggests their effects on our results is minimal because of our agreement with other contemporaneous measurements. Our close agreement with \cite{combi2019survey} is particularly significant, because their field of view encompasses much more of the coma than our observations and thus is not on the order of a collisional radius. Additionally, we think it is best to assume collisional equilibration in our calculations given that \cite{ootsubo2012akari} assumes the same, which allows us to make a comparison to their results in the following section that does not depend on the relationship chosen for expansion velocity.

\subsection{Inferred CO$_2$ abundances}
\label{disc-subsect:CO2-abundances}

Of the two calculated $\frac{CO_2}{H_2O}$ ratios for Lovejoy, the ratio inferred using the empirical release rates from \cite{mckay2015evolution} reproduces the direct measurement made by Spitzer within uncertainty, while the ratio inferred using the theoretical release rates from \cite{bhardwaj2012coupled} is approximately an order of magnitude lower than the direct measurement. The empirical release rates from \cite{mckay2015evolution} were calculated using data from a comet observed near solar maximum, C/2009 P1 (Garradd). Lovejoy was also observed near solar maximum in this work, and the empirical release rates were the most accurate at inferring a CO$_2$ abundance for Lovejoy using these observations. However, the empirical release rates were also the most accurate at inferring a CO$_2$ abundance for Lulin, which was observed near solar minimum. 

This could suggest that solar cycle phase plays little role and the discrepancy between the two sets of release rates is due to some other unknown factor. However, for infrared observations such as those from \cite{ootsubo2012akari}, the water production rate was determined from lines that could become optically thick for production rates on the order of $10^{29}~s^{-1}$ (such as theirs and those reported by other works cited in Section \ref{disc-subsubsect:lulin-h2o-prodrates}). If the coma was optically thick at the infrared observation wavelengths, the water production rate from \cite{ootsubo2012akari} could have been an underestimate. However, their CO$_2$ production rate was likely small enough to remain optically thin (on the order of $10^{27}~s^{-1}$), and was likely a more accurate estimate of the true production. Therefore, the $\frac{CO_2}{H_2O}$ ratio reported by \cite{ootsubo2012akari} could be an overestimate due to the H$_2$O production rate in the denominator being an underestimate. Ootsubo et al. addressed this possibility in an earlier work \citep{ootsubo2010detection} by measuring the water production rate via observations offset from the comet's optocenter. However, these measurements were taken about 6 weeks after our own, compared to the measurements from \cite{ootsubo2012akari} that are only 10 days before ours. Additionally, optocenter-derived vs. offset-derived production rate ratios may not be directly comparable due to asymmetries in the coma. Our measurements were taken centered on the optocenter, so our inferred $\frac{CO_2}{H_2O}$ ratio may not be comparable to the ratio reported in \cite{ootsubo2010detection}. For these reasons, comparison to the measurements from \cite{ootsubo2012akari} is likely still more relevant for our purposes. Using our own water production rate and the CO$_2$ production rate from \cite{ootsubo2012akari} to calculate the $\frac{CO_2}{H_2O}$ ratio could eliminate possible optical depth effects.

Even if the \cite{ootsubo2012akari} observations were not significantly affected by optical depth effects, measurements from the literature suggest a water production rate for Lulin that varies significantly with rotational phase. If the CO$_2$ production rate varies similarly, this may not be an issue for our purposes, since the $\frac{CO_2}{H_2O}$ ratio would remain unaffected. However, there are no data constraining how CO$_2$ production varies with rotational phase for Lulin, so we cannot know what effects this may or may not have on our results. Therefore, using our own water production rate to calculate the $\frac{CO_2}{H_2O}$ ratio is the best way we can attempt to eliminate rotational effects when comparing to our oxygen line ratio results, since at least the two measurements of water will then correspond to the same rotational phase.

We recalculated the $\frac{CO_2}{H_2O}$ reported in \cite{ootsubo2012akari} by dividing their CO$_2$ production rate by our water production rate. Using this updated ratio of $4.4\%~\pm~0.5\%$, the inferred abundance using the empirical release rates is no longer in agreement with the direct measurement. The abundance inferred using the theoretical release rates also does not accurately reproduce this updated ratio. However, several works using these release rates from \cite{bhardwaj2012coupled} report that they fail to reproduce direct measurements \citep{mckay2013observations,mckay2015evolution,mckay2016co}. Therefore, it is not surprising that using this set of release rates did not accurately reproduce the direct measurements for either of these comets. In contrast, the empirical release rates reproduced the direct measurement for Lovejoy, but not for Lulin (using the updated ratio from \cite{ootsubo2012akari}). 

Table \ref{disc-tab:comet-properties} contains key properties of these two comets, as well as two others that have rigorous oxygen line ratio analysis in the literature. Out of the four comets listed, Lulin is the outlier when examining the empirical release rates' accuracy at reproducing direct CO$_2$ measurements. While small differences are present between the comets' heliocentric distances, water production rates, and effective radii, Lulin is not a clear outlier for any of these properties. However, it is a clear outlier in both rotational phase and solar cycle phase. As discussed previously, we have attempted to account for rotational effects by using our own water production rate when calculating the $\frac{CO_2}{H_2O}$ ratio. Despite this, there are not enough observations of how gas production varies with rotational phase in Lulin to conclusively rule out rotational effects as a contributor to the discrepancy. However, if we assume our approach to the H$_2$O production rate eliminated any significant rotational effects, then the discrepancy in release rate accuracy could be explained by differences in solar activity assumptions between the empirical release rates and the actual observations. 

\begin{table*}
\caption{\label{disc-tab:comet-properties} Key properties that could explain the discrepancies in the effectiveness of the release rates for comets with oxygen line ratio analysis similar to ours. The first four data columns (dates observed, heliocentric distance, H$_2$O production rate, and solar cycle phase) all correspond to the following works that performed oxygen line ratio anaylsis with contemporaneous CO$_2$ comparison: Lovejoy and Lulin were observed with this work. C/2009 P1 (Garradd) was observed by \cite{mckay2015evolution}. C/2012 K1 (PanSTARRS) was observed by \cite{mckay2016co}. The remaining two columns (effective radius of nucleus and rotation period) contain values gathered from the literature, with reference information in the footnotes.}
\hspace*{-0.7cm}
\begin{tabular}{ccccccc}
\hline
Object & Date & $r$ & $Q$ & Solar cycle phase & Effective radius & Rot. period \\
& (UT) & (AU) & ($10^{29} s^{-1}$) & & (km) & (hr) \\
\hline
C/2014 Q2 (Lovejoy) & 2015 Feb 2-4 & 1.29 & 6.57 $\pm$ 0.74 & Max & 4.314 $\pm$ 0.227$^a$ & 17.89 $\pm$ 0.17$^b$ \\
C/2007 N3 (Lulin) & 2009 Feb 14-15 & 1.33 & 1.09 $\pm$ 0.11 & Min & 6.10 $\pm$ 0.25$^c$ & 41.45 $\pm$ 0.05$^d$ \\
C/2009 P1 (Garradd) & 2012 Mar 28-29 & 2.02 & 0.380 $\pm$ 0.038 & Max & 2.945$^a$-13.50$^c$ & 11.1 $\pm$ 0.8$^e$ \\
C/2012 K1 (PanSTARRS) & 2014 Jun 4 & 1.71 & 0.435 $\pm$ 0.044 & Max & 2.352$^a$-8.67$^f$ &9.2$^g$-9.4$^f$ \\
\hline
\end{tabular} \\
\raggedright
$^a$ \cite{paradowski2020new} \\
$^b$ \cite{serraRicart2015rotation} \\
$^c$ \cite{bauer2017debiasing,knight2024physical} \\
$^d$ \cite{bair2018coma} \\
$^e$ \cite{ivanova2017polarimetry} \\
$^f$ \cite{betzler2020bvr} \\
$^g$ \cite{mckay2016co} \\
\end{table*}

\subsection{Solar activity effects}
\label{disc-subsect:solar-activity-effects}

\begin{figure*}
    \centering
    \includegraphics[width=2\columnwidth]{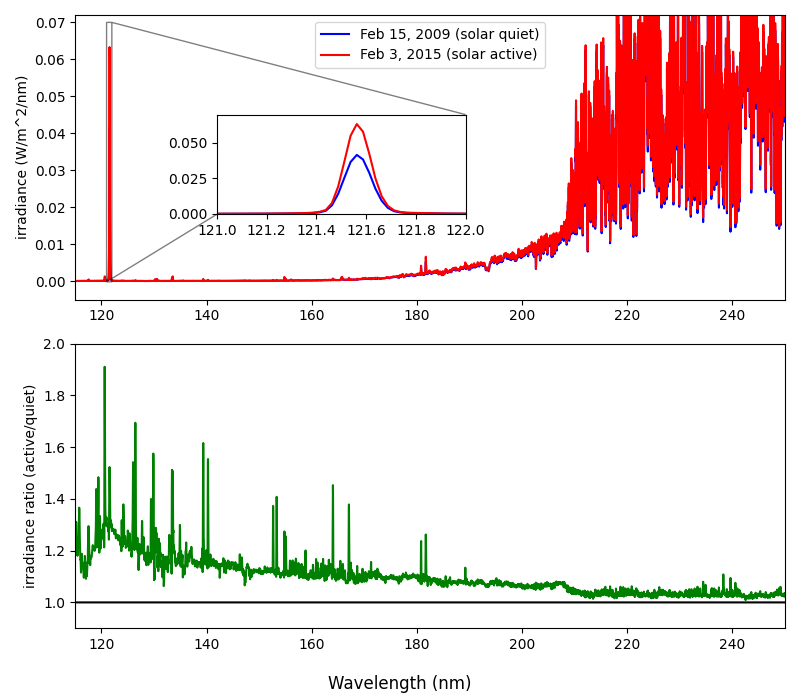}
    \caption{\label{disc-fig:lovejoy-lulin-solar-spectrum}Top panel: Total solar spectral irradiance for two dates plotted as a function of wavelength. The two dates plotted correspond to two of our observation dates, one each for Lovejoy and Lulin. The bottom panel plots the ratio between the two spectra (green), with the straight black line showing a ratio equal to one for easier visual comparison. The solar quiet spectrum is shown to have a consistently lower irradiance, such that the ratio is always greater than one. However, the difference is more pronounced at shorter wavelengths, particularly at the Lyman-alpha line, which is shown in a zoomed-in box in the top panel. Data retrieved from the LASP Interactive Solar Irradiance Datacenter (LISIRD) database using the NNL high-resolution model \citep{LISIRDdata}.}
\end{figure*}

Because oxygen atom release into the coma is caused by photodissociation via UV photons, changes in the sun's UV output could be reflected in a changing oxygen atom release rate. Indeed, previous work has also shown that solar activity affects the scale lengths of species such as water (\cite{cochran1993observational} and references therein). We retrieved the solar spectra on the days of our observations of both Lovejoy and Lulin from the LASP Interactive Solar Irradiance Datacenter (LISIRD) database \citep{LISIRDdata} in order to examine any differences between them. The top panel of Figure \ref{disc-fig:lovejoy-lulin-solar-spectrum} shows the total solar spectral irradiance plotted as a function of wavelength on two dates corresponding to our observations of Lovejoy (2015 February 3) and Lulin (2009 February 15). The bottom panel shows the ratio of the solar quiet spectrum to the solar active spectrum. The spectrum nearer to solar minimum (blue) maintains a consistently lower irradiance than the spectrum nearer to solar maximum. The bottom panel shows that the difference between the two spectra is more drastic at shorter wavelengths. It is likely this wavelength dependency that would be a more important driver of changes in the photochemistry, rather than the overall change in total flux. Indeed, the irradiance difference is especially noticeable at the Lyman-alpha line at $121.6~nm$, of which a zoomed-in view is shown in the top panel. Because the Lyman-alpha line is an important contributor to photodissociation rates, a difference in solar irradiance at this line due to the solar cycle could affect the photochemistry of [\ion{O}{1}] release, perhaps even enough to affect the inferred $\frac{CO_2}{H_2O}$ ratio from the oxygen line ratios. If so, it would be necessary to consider the solar cycle when deciding what release rates to use to calculate the oxygen line ratio. However, more work is needed to determine if the discrepancies we see in this work are consistent with what would be expected from a changing UV spectrum. This would likely require more detailed modeling that is beyond the scope of this work.

\section{Conclusion}
\label{sect:conclusion}

We present analysis of the forbidden oxygen lines at 5577, 6300, and 6364\AA~ for comets C/2014 Q2 (Lovejoy) and C/2007 N3 (Lulin). We calculate oxygen line ratios for each observation date and use them to calculate inferred CO$_2$ abundances, which we compare with direct measurements obtained by spacecraft near our observation dates. Our results suggest that the oxygen line ratio likely increases both with cometocentric and heliocentric distance. Additionally, we report water production rates for both comets that are consistent with most other values reported the literature when rotational variability and other effects such as optical depth are taken into consideration. When comparing CO$_2$ abundances, we find that using the empirical [\ion{O}{1}] release rates from \cite{mckay2015evolution} reproduces the direct CO$_2$ measurement for Lovejoy. When avoiding optical depth effects on water production in the coma at infrared wavelengths, we find that the empirical release rates do not reproduce the direct CO$_2$ measurement for Lulin. We suggest that this discrepancy could be due to differences caused by the solar cycle.

Because our inferred CO$_2$ abundance for Lovejoy using the empirical release rates accurately reproduced the direct measurement obtained by spacecraft, our results add to the growing body of work supporting the validity of the oxygen line ratio as a proxy for direct CO$_2$ measurement. However, the dependence of these inferred CO$_2$ abundances on the choice of release rates and the possible correlation between the release rates' accuracy and the solar cycle phase at the time of observation/release rate calculation suggests the factor of solar cycle phase needs to be incorporated into the oxygen line ratio method. Because of the strong possibility that the solar cycle plays a significant role in the photochemistry of [\ion{O}{1}] release, using the oxygen line ratio as a proxy for direct CO$_2$ measurement will not be a fully reliable method until this relationship is understood and quantified. The number of comets with rigorous [\ion{O}{1}] analysis that incorporates CO$_2$ abundance comparison is limited, and this work is to our knowledge the first to incorporate the solar cycle into this analysis. Future work should examine a more robust sample of comets observed at a variety of solar cycle phases to confirm and characterize the relationship between the solar cycle and release rate accuracy. Once this relationship's effect on coma photochemistry is more fully understood, the oxygen line ratio method may become an important strategy for determining CO$_2$ abundances, increasing our knowledge about this important molecule's behavior in cometary comae.

\begin{acknowledgements}
\label{sect:acknowledgements}
This work was supported by the NASA Solar System Workings Program through grant 80NSSC20K0140 and by NASA through an award issued by JPL/Caltech to Appalachian State University. This paper includes data obtained at the McDonald Observatory of The University of Texas at Austin. This paper also contains observations made with the Spitzer Space Telescope, which was operated by the Jet Propulsion Laboratory, California Institute of Technology under a contract with NASA. Support for the work with Spitzer was provided by NASA through an award issued by JPL/Caltech. We acknowledge the JPL Horizons System, which was used during observations to generate ephemerides for nonsidereal tracking of the comets. Additionally, we acknowledge the SIMBAD database, which was used for the selection of standard stars.
\end{acknowledgements}

\facility{Spitzer (IRAC), IRSA (Spitzer Heritage Archive)}
\facility{McDonald Observatory (2.7m Tull Coud\'e Spectrograph)}
\software{Scipy \citep{virtanen20-scipy}, Astropy \citep{astropy13,astropy18,astropy22}, sbpy \citep{sbpy19}, Pyspeckit \citep{ginsburg2022pyspeckit}}

\newpage
\bibliographystyle{aasjournalv7}
\bibliography{references.bib}{}

\end{document}